\documentstyle[12pt,epsfig]{article}
\begin{document}
\begin {center}
{\Large\bf Reply to Takamatsu's paper on the kappa}

\vskip 5mm

{D.V.~Bugg\footnote{email address: D.Bugg@rl.ac.uk}},   \\
{Queen Mary, University of London, London E1\,4NS, UK}
\vskip 2mm
\end {center}
\vskip 2.5mm
{\bf Abstract}
The essential elements in my way of fitting the $\kappa$ are
reviewed.
A number of incorrect claims made by Takamatsu are clarified and
corrected.
\section {Differences in approach}
The `sigma' group takes the $\kappa$ and $\sigma$ as conventional
Breit-Wigner resonances with widths linearly proportional to
phase space $\rho$ for decay.
These resonances alone do not fit the $K\pi$ (and $\pi \pi$) elastic
phase shifts.
An $s$-dependent background is added from a repulsive core so as to
achieve agreement with experiment.
The idea is described fully in Takamatsu's review.
This is one way of fitting the data, although the background needs to
be parametrised empirically.
The approach led to early clarification of the existence of the
$\kappa$ [2] and $\sigma$ resonances [3,4].

Since then, things have moved on.
My approach [5], and that of many others, is instead to treat the
$\kappa$ as a resonance with $s$-dependent width:
\begin {equation}
f_{el} = \frac {N(s)}{D(s)} =
\frac {M\Gamma _{K \pi}(s)} {M^2- s - iM\Gamma _{total}(s)}.
\end {equation}
No background is required.
In elastic $\pi \pi$ and $K\pi$ scattering, it is well known that
the Adler self-consistency condition, supplemented by linearity of
amplitudes within the Mandelstam triangle, leads to zeros at
$s = m^2_\pi/2$ for the $\pi \pi$ S-wave amplitude and at
$s = m^2_K - m_\pi^2/2$ for $K\pi$.
It accounts naturally for the fact that my $\Gamma (s)$ near threshold
is closely proportional to $(s - s_A)\rho$.
For the $\sigma$, data are more extensive and the required
$s$-dependence is somewhat more complicated: see Ref. [6] for
full details.

The denominator $D(s) $ in Eq. (1) originates from the right-hand cut
and should be common to elastic scattering and production.
Phases for the $\sigma$, derived from BES data for
$J/\psi \to \omega \pi ^+\pi ^-$, are indeed consistent with the same
$s$-dependence as $\pi \pi$ elastic scattering from 450 to 950 MeV
within experimental errors of $\sim 3.5^\circ$ [7].
Likewise $\kappa$ phases, derived from BES II data [8], are
consistent with the $s$-dependence of LASS data for $K\pi $ elastic
scattering [9] within similar errors.
These are important consistency checks which could fail if the initial
assumptions in Eq. (1) are wrong.

There is however more to the story.
The numerator $N(s)$ originates from the left-hand cut.
The `sigma' group makes no use of this information.
Neither did the first fits to production data.
This situation is now being remedied.
Caprini et al predict phase shifts for $\pi \pi$ elastic
scattering from the left-hand cut and impose consistency with
crossing symmetry using the Roy equation [10].
B\" uttiker et al carry through a similar calculation for $K\pi$
elastic scattering [11].
There are small discrepancies in masses and widths of $\sigma$ and
$\kappa$ fitted by individual groups and disagreements over some
details concerning effects of the $KK$ and $\eta \eta$ thresholds [6].
My own opinion is that those disagreements are within present errors.

In production processes, the left-hand cut is quite different to
elastic scattering, so there is no requirement for $N(s)$ to be the
same.
In $J/\Psi \to \omega \pi \pi$, left-hand singularities are very
distant, and it is not surprising that $N(s)$ is consistent with
a constant.
The same is true for the $\kappa$ in $J/\Psi \to K^*(890)\kappa$.
It is necessary to check whether any form factor is needed to
describe the distant singularity responsible for $N(s)$.
For the $\kappa$, any possible form factor is well determined [12] by
a combined fit to LASS, E791 [13] and BES II data.
The RMS radius is found to be $< 0.38$ fm with 95\% confidence.
Likewise for the $\sigma$, the data are consistent with a point
interaction; any possible form factor has if anything a marginally
negative RMS radius, but this is unphysical.

Takamoto, in Section 4 of his article, claims that I introduce
an {\it artificial suppression factor} between elastic
scattering and production reactions. This factor was, I
believe, first introduced by Au, Morgan and Pennington [14].
The discussion given above motivates its introduction.
The test is whether my approach (shared by many other groups)
fits all the data or not. With one small question mark discussed later,
it does.
The background in the approach of the `sigma' group needs to be
fitted empirically to each set of data.
The $s$-dependence of $\Gamma$ in my approach is of similar
flexibility to their background in a single reaction, but then
fits all sets of data self-consistently.
So, the approach based on Eq. (1) is more economical and has the
virtue of relating $N(s)$ of elastic scattering to the left-hand cuts.

In $\pi\pi$ elastic scattering, the left-hand cut is dominated
by $\rho$ exchange.
Formulae for this exchange are given by Zou and Bugg [15].
The $\rho$-exchange term is almost linear in $s$ and crosses zero
close to the Adler zero.
This is of course because the $\rho$ is a vector particle.
In $K\pi$ elastic scattering, $K^*$ exchange likewise dominates.

Takamatsu states that there is a {\it cancellation mechanism}
in the scattering amplitudes and that my relation between
elastic scattering and production {\it overlooks this cancellation
mechanism}.
His remark relates to the fact that, in the linear $\sigma$
model, a constant contact term cancels the amplitudes
from $s$, $t$ and $u$-channels $\sigma$ poles to produce the Adler zero
[16]. This contact term is making the non-$\rho$ exchange terms cancel
to zero at the Adler point.
I see no obvious reason why the same contact term should appear in
production processes which have drastically different left-hand cuts.

In the same section, Takamatsu refers to what he calls my
{\it so-called combined fit} to production and elastic data.
For $\pi \pi$, the fits are made [17] simultaneously
to BES II data on $J/\Psi \to \omega \pi ^+\pi ^-$ [18],
Cern-Munich elastic phase shifts [19], $K_{e4}$ data of Pisluk et al.
[20], four sets of data on $\pi \pi \to KK$ and one on
$\pi \pi \to \eta \eta$, and Kloe data on
$\phi \to \gamma \pi ^0 \pi ^0$ [21].
It is also constrained to fit as closely as possible to the phase shifts
predicted by Caprini et al. [10], so as to impose
consistency with left-hand cuts.
The maximum discrepancy with the phase shifts predicted by Caprini et
al. up to 750 MeV is $1.2^\circ$, i.e. about two standard
deviations. Fitting all these sets of data self-consistently is
non-trivial.

For the $\kappa$, my combined fits have been made [12] to LASS data on
$K\pi$ elastic scattering, E791 data on $D^+ \to (K^- \pi ^+ )\pi ^-$
and BES II data on $J/\psi \to K^+K ^-\pi ^+\pi ^-$ [8]; in this work,
the left-hand cut was not fitted.
Zhou and Zheng do fit the left-hand cut together with LASS data and
find a pole position of $694 \pm 53 - i(303 +- 30)$ MeV [22].
This compares with $750 ^{+30} _{-55} - i(342 +- 60) $ MeV from my
combined fit.
The discrepancy is within the errors.
Descotes-Genon and Moussallam find a $\kappa$ pole position of
$658 \pm 13 - i(279 \pm 12)$ MeV from their evaluation of the left-hand
cut [23].
This is somewhat lower and has small errors.
A question which needs attention here is the $s$-dependence of the
large sub-threshold contribution of $K\eta '$;
this could have a bearing on the apparent discrepancy, since coupling
of this channel to $\kappa$ and $K_0(1430)$ is not well separated
experimentally and affects what is fitted to the $\kappa$.
There are no direct data on the coupling of the $K\eta$ channel or
$\kappa \sigma$.
Consequently, fitting the $\kappa$ is subject to larger errors than
for the $\sigma$.
In my fits to LASS data [8], there is a  $2\sigma$ systematic
discrepancy around 1.2 GeV.
My suspicion is that this arises from
uncertainties in the treatment of the inelastic channels.

One remark is needed concerning fits made to the $\kappa$ by the `sigma'
group.
Their pole position is $841 \pm 30 ^{+81}_{-73} -i(309 \pm 90
^{+96}_{-144})$ MeV [24] but refers only to the Breit-Wigner part of
their amplitude; their background is treated as non-resonant.
For comparison with other determinations, which are all lower in mass,
it  is important to know the pole position from the coherent sums of
their  background plus Breit-Wigner resonances, i.e. from the full
S-wave amplitude.

\section {Observation of phase motion}
My paper on the $\kappa$ shows a determination of the $\kappa $ phase
in bins of $K\pi$ mass 100 MeV wide [8].
On p. 14, Takamatsu comments:
{\it It is a pity that there is no reference wave found with an adequate
background phase in the $\kappa$ region of the $K^*(892)K\pi$ channel.
The $K^*(1430)$ which interferes with the $\kappa$ is of no use,
since both amplitude and phase cannot be determined unambiguously
with an amplitude composed of [the] sum of two S-waves, even though the
amplitude and the phase of $K^*(1430)$ are determined by Breit-Wigner
parameters.}

There is no substance to this criticism.
Firstly, my paper states clearly [8] that
much of the phase determination of the $\kappa$ comes from its
interference with $K_1(1270)$ and $K_1(1400)$.
Secondly, the comment is wrong concerning interference between
$\kappa$ and $K_0(1430)$.

The $K_1 (1270)$ and $K_1(1400)$ occupy a $KK\pi \pi$ phase volume which
overlaps strongly with all masses of the $\kappa$. Masses and widths of
$K_1(1270)$ and $K_1(1400)$ may be separated in BES data via their
different couplings to $K^*\pi$ and $K\rho$. This is a point the
`sigma' group did not investigate, since they treated $K\rho$ as a
non-interfering `background'. Parameters of $K_1(1400)$ are consistent
with PDG values and are set to those values in my analysis. The
parameters of $K_1(1270)$ need a small adjustment from PDG values;
these parameters are essentially independent of $\kappa$ phases, since
they are fitted to the magnitudes of signals in the $K\rho$ and $K^*\pi$
channels.
When it comes to the bin-by-bin fit, parameters of $K_1(1400)$
and $K_1(1270)$ are allowed to vary within their errors; also
magnitudes and phases of all channels used in the global fit are set
free for every individual bin.
This allows the maximum possible freedom for every bin.
Feedback between the global fit and the bin-by-bin fit is small
($\sim 10\%$) and errors for
the $\kappa $ phases are covered generously by quoted errors.

Let us now turn to interferences between $K_0(1430)$ and $\kappa$.
This interference is effective over a limited
mass range, because of the width of $K_0(1430)$.
The $K_0(1430)$ mass and width are again determined from its
conspicuous line-shape in the overall $K\pi$ mass distribution.
As a matter of detail, a combined fit is made to BES and LASS data,
though the signal is larger in BES data.
Having assigned errors to $M$, $\Gamma$ of
$K_0(1430)$, the bin-by-bin fit determines the magnitude and phase of
the $\kappa$ in individual mass bins.
Feedback between the global fit
and the bin-by-bin fit is again $<10\%$.
So there is no truth in the assertion that the $\kappa$ amplitude and
phase cannot be determined independently.

My procedure is a conventional iterative process.
Similar procedures have been used to separate
$f_0(980)$ from $\sigma$ in $\pi \pi$ elastic scattering, and
$K_0(1430)$ from the broad $\kappa$ in LASS data, so there is nothing
new about this iterative process.
There is a built-in assumption that $K_0(1430)$ obeys analyticity
and hence has a Breit-Wigner shape.
If this were not true, one would see a systematic deviation in
$\kappa$ phases in the 1430 MeV mass range, but that does not arise.
If a background were required in addition to the $\kappa$ (or $\sigma$
in $\omega \pi \pi$ data), the phase in the bin-by-bin determination
would come out in systematic disagreement with the global fits. This
does not happen in my analyses. However, if the BES data are fitted to
the mass and width quoted by the `sigma' group, quite a large background
{\it is} required and is clearly an essential element in their fit.

\section {Fitting the $K^*(890)$ decay}
In the previous section, I have referred to the line-shape of
$K_0(1430)$.
In reality there is some $K_2(1430)$ present as well and I am able
to separate it.
This is an important technical point.

The $K_0(1430)$ and $K_2(1430)$ are formed largely in the
final states $K^*(890)K_0(1430)$ and $K^*(890)K_2(1430)$.
I include the decays of the $K^*$ in the partial wave analysis.
The BES analyses discard that information.
The angular correlations between
the $K^*$ decay and (a) the $K_J(1430)$ decay, (b) the production plane
of $K^*K_J$ depend in a very distinctive way on the spins $J$.
From the angular correlations, I can separate contributions made
by $K_0(1430)$, $K^*(1410)$ and $K_2(1430)$ cleanly.
The contribution from $K_0(1430)$ is in fact strongly dominant and
$K^*(1410)$ is insignificant - a large $K^*(1410)$ contribution is
unlikely because the $K^*(1410)$ has only a 6.6\% branching fraction
to $K\pi$.
The BES analysis of the $\kappa$ fits approximately equal
contributions from the three $K_J$ states, because they have no
information which separates them.
So in this respect, my analysis is definitely cleaner.

The two approaches are so divergent that it was impractical to
combine them in a single paper.
Takamoto says that I am not allowed to use the data.
I need to remind him that the final data set used by all sides
was produced by me and L.Y. Dong in January 2003, when my paper
was submitted to the collaboration.
The funding agency insisted on independent publication of my paper,
in view of the public money spent supporting this and other analyses
of BES data. I am grateful to the Royal Society for funding of this
work under contract Q772 with the Chinese Academy of Sciences.

\end{document}